\documentclass[%
 reprint,
 amsmath,amssymb,
 aps,
]{revtex4-2}

\usepackage{graphicx}
\usepackage{dcolumn}
\usepackage{bm}

\usepackage[utf8]{inputenc} 
\begin{document}


\title{A Unified Symmetry-Constrained Framework for Band Inversions in Photonic Crystals with $C_n$ Symmetry }

\author{Ze Tao}
\affiliation{Nanophotonics and Biophotonics Key Laboratory of Jilin Province, School of Physics, Changchun University of Science and
Technology, Changchun, 130022, P.R. China
}
\author{Fujun Liu}\email{Contact author: fjliu@cust.edu.cn}
\affiliation{Nanophotonics and Biophotonics Key Laboratory of Jilin Province, School of Physics, Changchun University of Science and
Technology, Changchun, 130022, P.R. China
}


\begin{abstract}
The lack of a unified theoretical framework for characterizing band inversions across different crystal symmetries hinders the rapid development of topological photonic band engineering. To address this issue, we have constructed a framework constrained by symmetry $k \cdot p$ that universally models bands near high-symmetry points for symmetric photonic crystals $C_6$, $C_4$, $C_3$, and $C_2$. This framework enables a coefficient-free quantitative diagnosis of band topology. We have demonstrated the power of this framework by systematically engineering band inversions. In $C_6$ crystals, we induce a reopening of the linear gap at $\Gamma$. In $C_4$ systems, mirror symmetry enforces a characteristic quadratic coupling leading to distinct spectral features. Our analysis further reveals that a lone $E$ doublet prevents inversion at the $\Gamma$ point in $C_3$ symmetry, while $C_2$ symmetry facilitates a unique inversion of $Y$ pointsints with anisotropic gap. This symmetry-first, fit-free approach establishes a direct link between experimental band maps and the extraction of fundamental topological parameters. It offers a universal tool for inversion and coupling-order identification.  
\end{abstract}

\maketitle

\textit{Introduction.-}
Topological photonic bands and inversion engineering open a key route to robust modal control and programmable coupling in integrated optics and metamaterials, with clear physical interpretability and broad application prospects\cite{yan2025realization,chen2025one,love2025programmable,vaidya2023topological,camara2025symmetry}. A particularly critical aspect of this engineering lies in understanding transverse magnetic (TM) bands near high symmetry points, where point-group-governed coupling orders and anisotropy directly determine gap closing/reopening behavior and topological transitions\cite{wang2023hybrid,xiong2024scattering,ni2024three,camara2025symmetry}. However, progress in this domain has been hampered by the absence of a unified theoretical framework. We therefore recognize the urgent need for a symmetry-first, portable $k \cdot p$ model that simultaneously covers $C_6$, $C_4$, $C_3$, and $C_2$ crystals while interfacing seamlessly with computational outputs like projectors, dispersions, and Brillouin-zone maps.

Prior studies have made significant advances through representation-theoretic and $k\!\cdot \!p$ tight-binding paradigms, including identification of band inversions, selection rules, and coupling orders under specific point groups\cite{jia2025diagnosis,yin2023topological,gao2023topological,chen2024realization}, however, these approaches remain fundamentally limited by their specificity. Current methods are often siloed within single-symmetry groups, frequently rely on empirical coefficient fitting, and consequently struggle to directly extract fundamental parameters like the mass term, minimal coupling order, and anisotropy from standard computational data\cite{istrate2006photonic,wu2015scheme,ochiai2009photonic,chen2017valley,zhu2018topological,tserkezis2010retrieving}. This limitation primarily stems from a widespread oversight of the joint structure between partnered doublets and low-energy subspaces, which precludes a unified treatment of crucial features like mirror-aligned anisotropy and direction-dependent slope variations across different symmetries.

To overcome these limitations, we propose a symmetry-constrained unified $k \cdot p$ model specifically designed to model TM bands near high-symmetry points through two low-energy subspaces and a partnered doublet. This framework provides a universal model spanning $C_6$, $C_4$, $C_3$, and $C_2$ symmetries that enables the direct, fit-free extraction of the mass term and minimal coupling order from projectors, dispersions, and Brillouin-zone maps. By properly accounting for the relationship between doublets and subspaces, our approach simultaneously resolves mirror-aligned anisotropy, thereby establishing a reusable pipeline for quantitative parameter extraction across multiple point groups.

We demonstrate the power and versatility of this framework through comprehensive analysis of all four symmetries. For $C_6$ systems, we show how intracell contraction drives the $\Gamma$-point mass through zero, enabling precise quantification of the linear reopening governed by first-order selection rules. In $C_4$ crystals, we leverage mirror symmetries to suppress linear mixing and enforce quadratic coupling, resulting in characteristic spectral features including an \textbf{X}-shaped zero set in $\omega_4 - \omega_3$ and a $\Gamma$-centered dome in $\omega_2 - \omega_1$ without inversion. For $C_3$ and $C_2$ symmetries, we explain why a lone $E$ doublet prevents $\Gamma$-point inversion and demonstrate how anti-diagonal dimerization instead produces a $Y$-point inversion with distinctly unequal linear reopenings along $Y \to \Gamma$ and $Y \to M$ directions. Through this systematic investigation, we close a significant methodological gap in the field by delivering a symmetry-first, fit-free, and reusable pipeline for quantitative band structure analysis across point groups.

\textit{General Theory of Symmetry-Governed Coupling in Photonic $k \cdot p$ Models.-}We analyze the $k\cdot p$ expansion of the two-dimensional TM Maxwell eigenproblem in a neighborhood of the $\Gamma$ point. Throughout this work, we operate within the 2D TM sector (\(\mu=1\)) with Bloch-periodic fields, casting the underlying Maxwell master equation as a Hermitian generalized eigenproblem $\mathcal{A}\,\psi=\frac{\omega^{2}}{c^{2}}\,\mathcal{B}\,\psi,$ with \(\mathcal{A}=\nabla\times\nabla\times\) and \(\mathcal{B}=\varepsilon(\mathbf r)\) in curl–curl form. Equivalently, in the scalar TM reduction, this becomes:
\begin{equation}
-\nabla\!\cdot\!\big[\varepsilon(\mathbf r)^{-1}\nabla \psi\big]=\frac{\omega^{2}}{c^{2}}\,\psi .
\end{equation}
This continuum formulation serves as the foundation for our $k \cdot p$ model, with detailed implementation provided in the Supplementary Materials (SM). 

Our approach models the low-energy sector using two two-dimensional subspaces: one labeled $p$ (corresponding to the numerically identified "$p$-type" doublet) and another labeled $d$ (matching the "$d$-type" doublet). We tie the basis and projection methodology directly to our computational implementation: for $C_6$ and $C_4$ calculations, we define the $p/d$ subspaces through script-built projectors acting on numerical eigenfields, distinguishing two partner doublets based on these projections; for $C_3$ calculations, we adopt the $E_\pm$ projection across three equivalent sites to define a single partner doublet; and for $C_2$ calculations, we exploit \(C_2\) even/odd symmetry to extract at \(\Gamma\) a near-degenerate partner doublet and supplement it with an analytical \(p\!\!-\!\!d\) subspace projection. We construct the effective model from the abstract constituents—two two-dimensional subspaces (\(p\) and \(d\)) together with a partner doublet—and, for each point group, we impose specific projection rules that precisely fix these subspaces and their pairing. We document in the Supplementary Materials (SM) the technical details of intra-subspace bases (including circular bases) and the corresponding gauge-equivalent formulations.

Numerical implementations across all four symmetries reveal two two-dimensional low-energy subspaces ($p$ and $d$) near the $\Gamma$ point. We work in the orbital subspace spanned by \(\{p,d\}\) and define
$
\tau_z := P_d - P_p,$ $
\tau_x + i\tau_y := 2\,\Pi_{dp},
$
where $P_{p,d}$ denote projectors onto the $p$ and $d$ subspaces, and $\Pi_{dp}$ represents the transition operator from $p$ to $d$. In any orthonormal ${p,d}$ basis, these definitions reduce to standard $2\times2$ Pauli matrices. The computational frameworks organize states into two partnered blocks labeled by $\eta=\pm$: $C_6/C_4$ utilize $p/d$ projection, $C_3$ employs $E_\pm$ projection, and $C_2$ uses $C_2$ even/odd pairing at $\Gamma$ combined with analytical $p/d$ subspace projection. For each block $\eta$, we express the minimal effective Hamiltonian as:
\begin{equation}\label{1}
    h_\eta(\mathbf{k}) = d_0(\mathbf{k}) \, \tau_0 
+ d_z(\mathbf{k}) \, \tau_z 
+ \left[ \Phi_\eta(\mathbf{k}) \, \tau_+ + \Phi_\eta^*(\mathbf{k}) \, \tau_- \right],
\end{equation}
where $\tau_0$ is the $2\times2$ identity matrix and $\tau_\pm := \tfrac{1}{2}(\tau_x \pm i\tau_y)$. The components are defined as follows: (i) $d_0(\mathbf{k})$ represents a scalar even function (to second order, $C_x k_x^2 + C_y k_y^2$) that shifts the band center without affecting the gap; (ii) $d_z(\mathbf{k}) = M + \mathbf{k}^\top B \mathbf{k} + O(k^4)$, with $B$ a real symmetric $2\times2$ tensor, where we take the $\Gamma$-point center frequencies \(\omega_p(0)\) and \(\omega_d(0)\) of the \(p\) and \(d\) doublets and set \(M=\tfrac{1}{2}\big(\omega_d(0)-\omega_p(0)\big)\), obtained through code projections; (iii) Point-group symmetry and the specified partner pairing determine the lowest-order inter-orbital invariant \(\Phi_\eta(\mathbf{k})\), which obeys:
\begin{equation}\label{2}
\Phi_\eta(R_\theta \mathbf{k}) = e^{i m \theta}\,\Phi_\eta(\mathbf{k}),
\qquad
H(\mathbf{k}) = H^*(-\mathbf{k}),
\end{equation}
where we use \(R_\theta\) for a \(C_n\) rotation and define \(m\) as the minimal allowed crystal–angular-momentum order determined by the \(p\)–\(d\) representation difference at \(\Gamma\) together with the pairing rule. We write the band energies as:
\begin{equation}\label{3}
\omega_{\eta,\pm}(\mathbf{k}) = d_0(\mathbf{k}) \pm \sqrt{d_z(\mathbf{k})^2 + \big|\Phi_\eta(\mathbf{k})\big|^2}\, .
\end{equation}
Eqs.~\eqref{1}–\eqref{3} maintain gauge invariance under any unitary transformation within ${p,d}$, as such transformations only shift the phase of $\Phi_\eta$ between $\tau_x$ and $\tau_y$ while preserving $\omega_{\eta,\pm}$ and all observables. The Supplementary Materials (SM) provide detailed specifications of point-group–specific minimal $\Phi_\eta$ forms (linear for $C_6/C_3/C_2$; quadratic for $C_4$ when mirrors are preserved), mirror-line conditions for $\Phi_\eta\equiv 0$, and interface reduction procedures ($k_y\to -i\partial_y$) with corresponding criteria. Therefore, we employ Eqs.~\eqref{1}–\eqref{3} as our fundamental framework while deliberately avoiding coefficient fitting. Instead, we directly extract information from computational outputs: (i) we determine the sign and inversion of $M$ at $\Gamma$ from partnered sets and projections ($p/d$ for $C_6/C_4$, $E_\pm$ for $C_3$, $C_2$ even/odd plus analytical $p/d$ for $C_2$); (ii) we validate for $C_4$ that the zero contour of $\Delta_{34}(\mathbf{k})$ from full Brillouin zone scans aligns with mirror lines, forcing $\Phi_\eta(\mathbf{k})\equiv 0$ and confirming $m=2$; (iii) we corroborate for $C_2$ the generic linear inter-orbital coupling using the $\Gamma$ even/odd basis and anisotropic dispersions/projections along two high-symmetry lines. We keep $B$ and $(C_x, C_y)$ symbolic and focus on gauge-invariant diagnostics. We perform all calculations in the two-dimensional TM sector under Bloch-periodic boundary conditions. Using finite-difference Bloch implementations for \(C_4\) and \(C_2\), we assemble the Hermitian generalized eigenvalue problem \(A\psi=\lambda B\psi\) and read off frequencies via \(\omega=\sqrt{\lambda}\). We impose Bloch-phase factors on the five-point stencil when crossing unit-cell interfaces. For $C_2$ at $\Gamma$, projections and orthogonalization employ the $B$-metric inner product, while $C_4$ calculations compute $p$/$d$ projections as real-space cell integrals against analytic templates without invoking the $B$-metric. Spectral/plane-wave expansion implementations ($C_6, C_3$) assemble inverse-dielectric Fourier matrices and solve standard Hermitian eigenproblems in truncated $\mathbf{G}$ bases, with band plots following hard-coded high-symmetry paths. We fix units by setting \(a=1\) and \(c=1\), which yields \(\omega \propto a^{-1}\); we address absolute scaling in the Supplementary Materials (SM).

We report main-text diagnostics taken directly from the script outputs and refrain from any parameter fitting. At $\Gamma$, partner ordering combined with specified projections ($p$/$d$ for $C_6$/$C_4$, $E\pm$ (three-site sampling) for $C_3$, and $C_2$ even/odd plus analytical $p$/$d$ projection for $C_2$) determines the mass sign $M$ and band inversion occurrence. For $C_4$, Brillouin-zone scans produce $\Delta_{34}(\mathbf{k})$ maps whose zero contours coinciding with mirror-invariant lines provide direct evidence that $\Phi\eta(\mathbf{k}) \equiv 0$ on these lines, confirming quadratic minimal order ($m = 2$). Within \(C_6\), a topologically trivial supercell spectrum exhibits two counter-propagating in-gap branches; we assign colors according to edge-localization scores (normalized interface-region energy-density integrals) that diagnose the \(m=1\) edge character. For $C_2$, the $\Gamma$ even/odd basis and projections along high-symmetry lines corroborate generic linear inter-orbital coupling while revealing tiny avoided crossings away from $\Gamma$. We keep $B$ and $(C_x, C_y)$ symbolic and focus on gauge-invariant diagnostics---$M$, $m$, and mirror-line zeros.  

\textit{$C_{6}$ Hexagonal Cluster: $\Gamma$-point inversion under intra-cell contraction.-}\begin{figure*}[t] 
\centering
  \includegraphics[width=\textwidth]{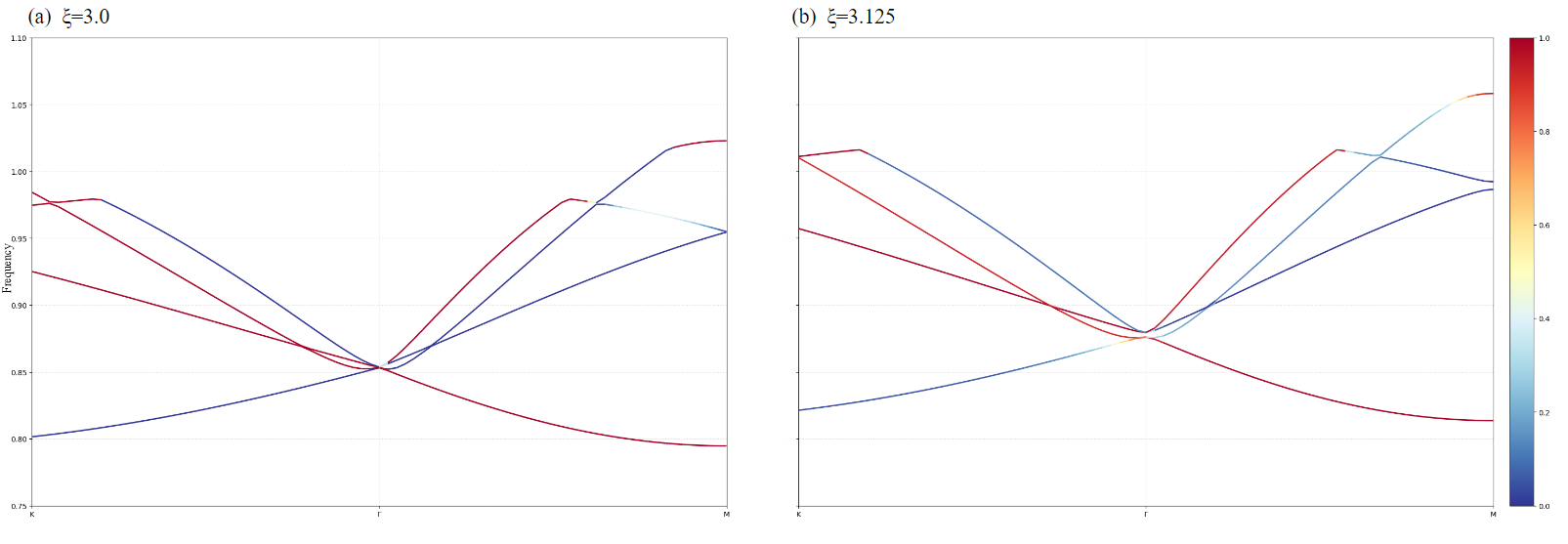} 
\caption{$C_6$ hexagonal cluster (spectral/PWE): $\Gamma$-point inversion under intra-cell contraction. Bulk TM bands along $K$–$\Gamma$–$M$ (units $\omega a/2\pi c$); we map colors to the normalized \(p_+\) weight \( |p_+|^2/(|p_+|^2+|d_+|^2) \) (blue \(p\)-like, red \(d\)-like). We tune \(\xi \equiv a_0/R_{\text{hex}}\) (larger \(\xi\) pulls pillars inward; pillar radius scales with \(R_{\text{hex}}\)). Near \(\Gamma\) we use the linear \(m=1\) block \(h_\pm(\mathbf{k})=\big(M+\mathbf{k}^\top B\,\mathbf{k}\big)\tau_z + A\,(k_x\tau_x \mp k_y\tau_y)\), fixed by the selection rule \(\Delta\ell=1\) between \(p\) and \(d\) doublets. (a) \(\xi=3.0\): the two colored branches touch at \(\Gamma\), exchange \(p/d\) character, diagnose \(M\simeq 0\), and reopen linearly away from \(\Gamma\). (b) \(\xi=3.125\): we see an inverted ordering at \(\Gamma\) (lower \(d\)-dominant, upper \(p\)-dominant), hence \(M<0\); dispersions and characters vary smoothly along \(K\)–\(\Gamma\) and \(\Gamma\)–\(M\). The mass follows \(M=\tfrac12[\omega_d(0)-\omega_p(0)]\) from the same projectors. Reciprocal-lattice convention: \(\mathbf{a}_1=(a,0)\), \(\mathbf{a}_2=(a/2,\sqrt{3}\,a/2)\); \(\mathbf{b}_1=(2\pi/a,-2\pi/(\sqrt{3}\,a))\), \(\mathbf{b}_2=(0,4\pi/(\sqrt{3}\,a))\); \(\Gamma=(0,0)\), \(K=(4\pi/(3a),0)\), \(M=(\pi/a,\pi/(\sqrt{3}\,a))\); symmetry-related points are equivalent.}
  \label{C6}
\end{figure*}
We investigate a hexagonal artificial-atom cluster on a triangular lattice using a spectral/plane-wave expansion (PWE) solver along the $K$-$\Gamma$-$M$ high-symmetry path. We quote frequencies in the dimensionless unit \(\omega a/(2\pi c)\) with the convention \(a=1\) and \(c=1\). The control parameter is a dimensionless intra-cell contraction $\xi$, defined as the ratio of the lattice constant to the cluster radius; increasing $\xi$ pulls the six pillars inward while maintaining a fixed ratio between pillar radius and cluster radius. Colored traces represent the normalized $p_+$ weight $|p_+|^2/(|p_+|^2 + |d_+|^2)$, where blue indicates predominantly $p$-like character and red indicates $d$-like character.

Within the general block Hamiltonian framework $h_\eta(\mathbf{k})$ established in Eqs.~\eqref{1}–\eqref{3}, We define \(M=\tfrac{1}{2}\big(\omega_d(0)-\omega_p(0)\big)\) as half the \(\Gamma\)-point splitting and take \(\omega_{p}(0)\), \(\omega_{d}(0)\) as the \(\Gamma\)-point center frequencies of the two doublets, which we extract using the same projectors that generate the orbital-color mapping. For $C_6$ symmetry, the minimal crystal-angular-momentum order in the inter-orbital invariant is $m = 1$. This arises because the $p$ and $d$ doublets transform with angular momenta $\ell_p = \pm 1$ and $\ell_d = \pm 2$ (modulo 6), respectively, while $k_\pm$ carry $\pm 1$ angular momentum. The product $p_\pm^\dagger d_\pm k_\mp$ is rotationally scalar, permitting a linear invariant that connects $p$ and $d$ orbitals. The resulting $k \cdot p$ block Hamiltonian takes the form $h_\pm(\mathbf{k}) = \big( M + \mathbf{k}^\top B \mathbf{k} \big) \tau_z + A , (k_x \tau_x \mp k_y \tau_y)$, with corresponding dispersion relation $\omega_\pm(\mathbf{k}) = d_0(\mathbf{k}) \pm \sqrt{\big( M + \mathbf{k}^\top B \mathbf{k} \big)^2 + |A|^2 k^2}$.

Fig.~\ref{C6} presents bulk-band results for two distinct intra-cell contraction parameters. For $\xi=3.0$, we observe two colored branches that touch tangentially at $\Gamma$ and exchange their $p/d$ orbital components, realizing the model's critical point with $M\simeq 0$. Away from $\Gamma$, the gap grows linearly with $|\mathbf{k}|$, with the coefficient $|A|$ controlling the slope—consistent with the $C_6$ case where the representation difference $\Delta\ell=1$ fixes the minimal order at $m=1$ for linear coupling. When we increase $\xi$ to $3.125$, a clear band-order inversion occurs at $\Gamma$: the lower branch becomes $d$-dominated while the upper branch becomes $p$-dominated, indicating $M<0$. Along both the $K$–$\Gamma$ and $\Gamma$–$M$ paths, the dispersion relations and orbital compositions remain continuous and smooth.

These results demonstrate that increasing intra-cell contraction (larger $\xi$) monotonically drives the mass term $M$ from positive values through zero to negative values, while the selection rule maintains the lowest-order invariant at $m=1$. We demonstrate that varying \(\xi\) serves as a control parameter for \(M\), and symmetry uniquely fixes the fundamental coupling order.

\textit{$C_4$ Square Cell: quadratic $m=2$ coupling and mirror-aligned anisotropy.-}\begin{figure*}[t] 
  \centering
   \includegraphics[width=\textwidth]{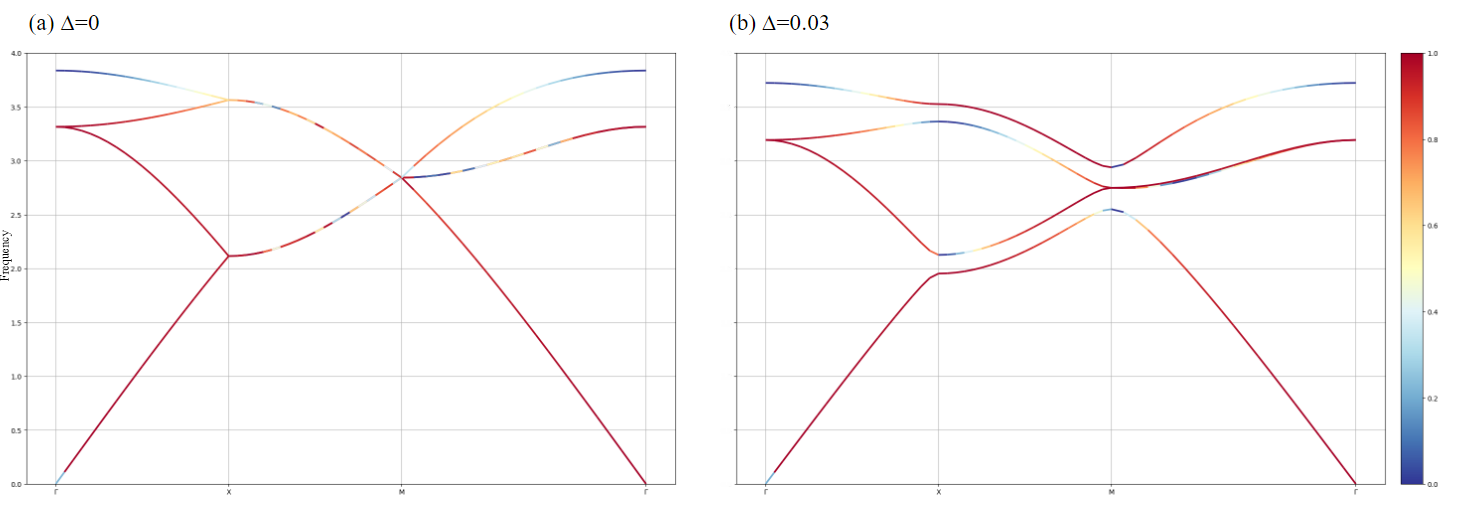}
\caption{$C_4$ square cell (FD--Bloch): TM bands along $\Gamma \to X \to M \to \Gamma$ at two inward shifts $\Delta$. We compute a square unit cell with four Si pillars pulled inward along the diagonals and color the branches through the normalized $p_+$ projector $|p_+|^2/(|p_+|^2 + |d_+|^2)$ (blue $p$-like, red $d$-like).  
(a) $\Delta = 0$: the partner pair near $\Gamma$ opens quadratically in $|\mathbf{k}|$ and keeps its ordering. (b) $\Delta = 0.03$: stronger contraction increases curvature and separation without producing a $\Gamma$-centered inversion. We diagnose an \(m=2\) block
\(h_\pm(\mathbf{k})=\big(M+\mathbf{k}^\top B\,\mathbf{k}\big)\tau_z+\alpha\,(k_x^2-k_y^2)\tau_x \pm \beta\,(2k_xk_y)\tau_y\), with \(C_{4v}\) mirrors fixing the coupling structure. We set \(\mathbf{a}_1=(a,0)\), \(\mathbf{a}_2=(0,a)\) and \(\Gamma=(0,0)\), \(X=(\pi/a,0)\), \(M=(\pi/a,\pi/a)\), and we measure frequencies in \(\omega a/(2\pi c)\). }
  \label{C41}
\end{figure*}
\begin{figure*}[t] 
  \centering
   \includegraphics[width=\textwidth]{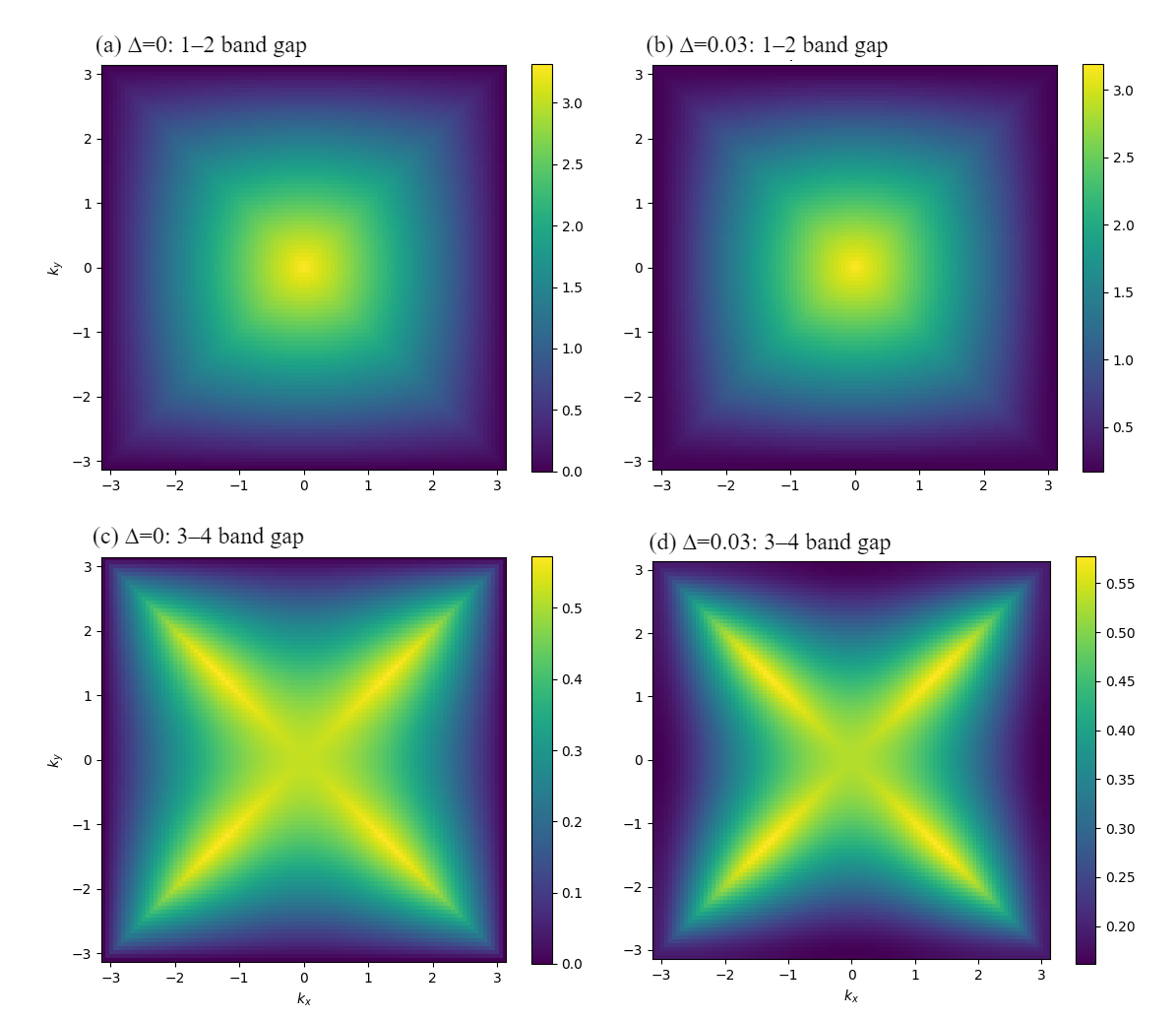} 
  \caption{$C_{4}$ square cell (FD--Bloch): two-dimensional Brillouin-zone gap maps at $\Delta = 0$ and $\Delta = 0.03$. Top row plots $\omega_2 - \omega_1$; the gap forms a nearly isotropic dome centered at $\Gamma$ and decays toward the zone edge. Bottom row plots $\omega_4 - \omega_3$; the gap draws an X-shaped ridge aligned with crystal mirrors---maxima along $k_x = \pm k_y$ and minima along $k_x = 0$ or $k_y = 0$. Increasing $\Delta$ mainly rescales the magnitude while preserving this geometry. The angular pattern follows directly from the quadratic invariants $(k_x^2 - k_y^2,\, 2 k_x k_y)$ in the $m = 2$ $k \cdot p$ block.}
  \label{C42}
\end{figure*}We model a square unit cell in which four silicon pillars move radially inward along the diagonals, using the dimensionless control parameter \(\Delta\) to set the displacement amplitude (where $\Delta = 0$ represents the baseline configuration and $\Delta = 0.03$ indicates stronger contraction). We compute TM bands with a finite-difference Bloch solver along \(\Gamma\!\to\!X\!\to\!M\!\to\!\Gamma\) and color each branch by the normalized \(p_+\) projector weight \( |p_+|^2/(|p_+|^2+|d_+|^2) \). We set the lattice vectors \(\mathbf{a}_1=(a,0)\), \(\mathbf{a}_2=(0,a)\), derive the reciprocal vectors \(\mathbf{b}_1=(2\pi/a,0)\), \(\mathbf{b}_2=(0,2\pi/a)\), and specify the high-symmetry points \(\Gamma=(0,0)\), \(X=(\pi/a,0)\), \(M=(\pi/a,\pi/a)\); we present the band path in Fig.~\ref{C41} and display the two-dimensional Brillouin-zone gap maps in Fig.~\ref{C42}.

The square cell preserves the mirror symmetries of the $C_{4v}$ point group. These mirror operations forbid linear inter-orbital coupling in the $p$/$d$ sector, consequently forcing the lowest symmetry-allowed coupling to appear at quadratic order. Within the framework of Eqs.~\eqref{1}--\eqref{3}, we express the Hamiltonian as $h_\pm(\mathbf{k}) = \big( M + \mathbf{k}^\top B \mathbf{k} \big) \tau_z + \alpha , (k_x^2 - k_y^2) \tau_x \pm \beta , (2 k_x k_y) \tau_y,$ which corresponds to $m = 2$ coupling. Along the $\Gamma$--$X$ direction ($k_y = 0$), only the $\alpha$ term contributes; along $\Gamma$--$M$ ($k_x = k_y$), both invariants combine to enhance the splitting. This model predicts both a quadratic gap opening near $\Gamma$ and a mirror-aligned angular anisotropy. Fig.~\ref{C41}(a) ($\Delta = 0$) maintains the band ordering near $\Gamma$ and demonstrates the characteristic $|\mathbf{k}|^2$ opening, while Fig.~\ref{C41}(b) ($\Delta = 0.03$) shows strengthened curvature and separation without inducing a $\Gamma$-centered inversion.

The Brillouin-zone gap maps further illustrate these symmetry-imposed features. Figs~\ref{C42}(a) and~\ref{C42}(b) display $\omega_2 - \omega_1$ at $\Delta = 0$ and $\Delta = 0.03$, respectively, revealing a nearly isotropic dome centered at $\Gamma$ that decays toward the zone edge. In contrast, Figs.~\ref{C42}(c) and~\ref{C42}(d) plot $\omega_4 - \omega_3$, showing an X-shaped ridge structure aligned with the crystal mirrors—exhibiting maxima along $k_x = \pm k_y$ and minima along $k_x = 0$ or $k_y = 0$. Increasing $\Delta$ preserves these fundamental geometric patterns while primarily rescaling the magnitude.  

\textit{$C_3$ Triangular Cell: no $\Gamma$-point inversion; $E_\pm$ tracking under intra-cell contraction.-}
\begin{figure*}[t]
  \centering
    \includegraphics[width=\textwidth]{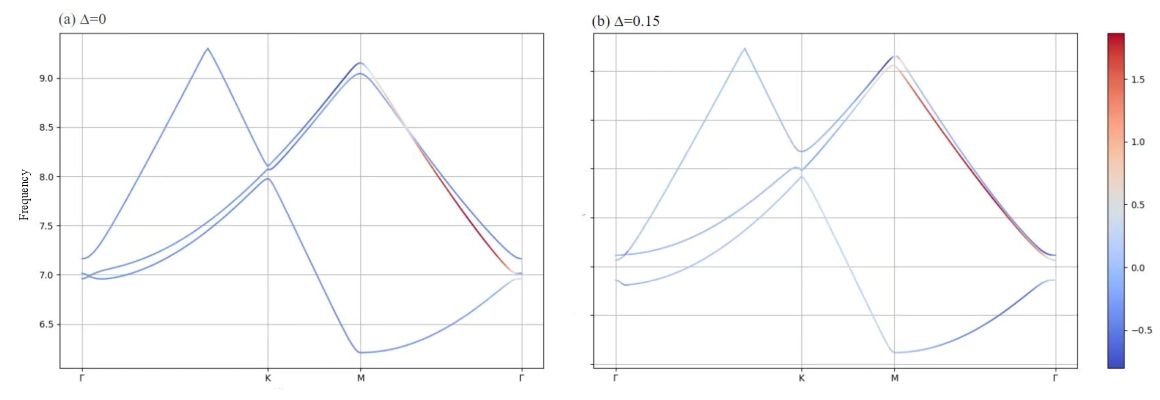}
  \caption{C$_3$ triangular cell (spectral/PWE, TM). Bands along $\Gamma\!\to\!K\!\to\!M\!\to\!\Gamma$ at two intra-cell contractions: (a) $\Delta=0$, (b) $\Delta=0.15$. Three identical Si pillars occupy the vertices of an equilateral triangle at polar angles $0^\circ,120^\circ,240^\circ$ on a ring of radius $R-\Delta$ (we keep a fixed pillar-to-ring radius ratio). Colors indicate the group-theoretic $E_\pm$ character from three-site sampling: red favors $E_+$ and blue favors $E_-$. Both panels show no $\Gamma$-centered inversion and no swap of $E_\pm$ character at $\Gamma$.}
  \label{fig:C3}
\end{figure*}
We investigate a triangular Bravais lattice with three identical pillars, controlling their configuration through a dimensionless contraction parameter $\Delta$ that radially displaces sites toward the cell center. We apply a spectral plane-wave–expansion (PWE) scheme to the TM Maxwell eigenproblem and evaluate along \(\Gamma\!\to\!K\!\to\!M\!\to\!\Gamma\). We define the lattice vectors as $\mathbf a_1=(a,0)$ and $\mathbf a_2=(a/2,\sqrt3,a/2)$, with corresponding reciprocal vectors $\mathbf b_1=(2\pi/a,,-2\pi/(\sqrt3 a))$ and $\mathbf b_2=(0,,4\pi/(\sqrt3 a))$; we place the high-symmetry points at \(\Gamma=(0,0)\), \(K=\big(4\pi/(3a),\,0\big)\), and \(M=\big(\pi/a,\,\pi/(\sqrt{3}\,a)\big)\). To track the $E$ doublet behavior, we employ standard three-site projectors $S_{E_\pm}$ and color bands according to the difference $S_{E_+}-S_{E_-}$.

Fig.~\ref{fig:C3} presents our computational results. For both $\Delta=0$ and $\Delta=0.15$, the spectrum near $\Gamma$ contains only a single $E$ doublet that remains well-separated from other states. The $E_\pm$ weights evolve smoothly along both $K\!-\!M$ and $M\!-\!\Gamma$ paths without exhibiting any exchange behavior at $\Gamma$. Within the unified model framework established in Eqs.~\eqref{1}–\eqref{3}, $C_3$ symmetry does permit a linear inter-orbital invariant ($m=1$); however, this particular geometry does not produce a second near-resonant subspace at $\Gamma$ that could form a partnered pair and drive $M\!\to\!0$. Consequently, we observe no $\Gamma$-point band inversion across the entire scanned range of $\Delta$ values.

To model the $\Gamma$-neighborhood behavior, we consider two subspaces that would form a partnered pair if a second $E$-type doublet approached $\Gamma$. $C_3$ symmetry fixes the minimal angular order at $m=1$, leading us to retain the linear inter-orbital invariant and express the block Hamiltonian as:
$h_\pm(\mathbf k)
  = \big(M+\mathbf k^{\mathsf T}B\,\mathbf k\big)\,\tau_z
  + A\,(k_x \tau_x \mp k_y \tau_y),$
where $\tau_{x,y,z}$ represent Pauli matrices acting on the partnered ${p,d}$-type subspace (with identity $\tau_0$), $B$ denotes a real symmetric $2\times2$ tensor, and $A$ is a real coupling constant. Equivalently, we can express $\Phi_\pm(\mathbf k)=A,k_{\mp}$ with $k_\pm=k_x\pm i k_y$ and $\tau_\pm=(\tau_x\pm i\tau_y)/2$. The resulting dispersion relation reads: 
$\omega_{\pm}(\mathbf k)
  = d_0(\mathbf k)\ \pm\
  \sqrt{\big(M+\mathbf k^{\mathsf T}B\,\mathbf k\big)^2 + A^2 k^2 } .$

In this triangular configuration, however, we consistently observe only one $E$ doublet near $\Gamma$ throughout the $\Delta=0\to0.15$ range. The effective mass $M$ therefore remains detuned from zero, preventing the linear $m=1$ coupling from triggering a $\Gamma$-point inversion and resulting in no swapping of $E_\pm$ character at $\Gamma$.

\textit{$C_2$ Square Cell: Y-point inversion under anti-diagonal dimerization.-}
\begin{figure*}[t]
\centering
  \includegraphics[width=\textwidth]{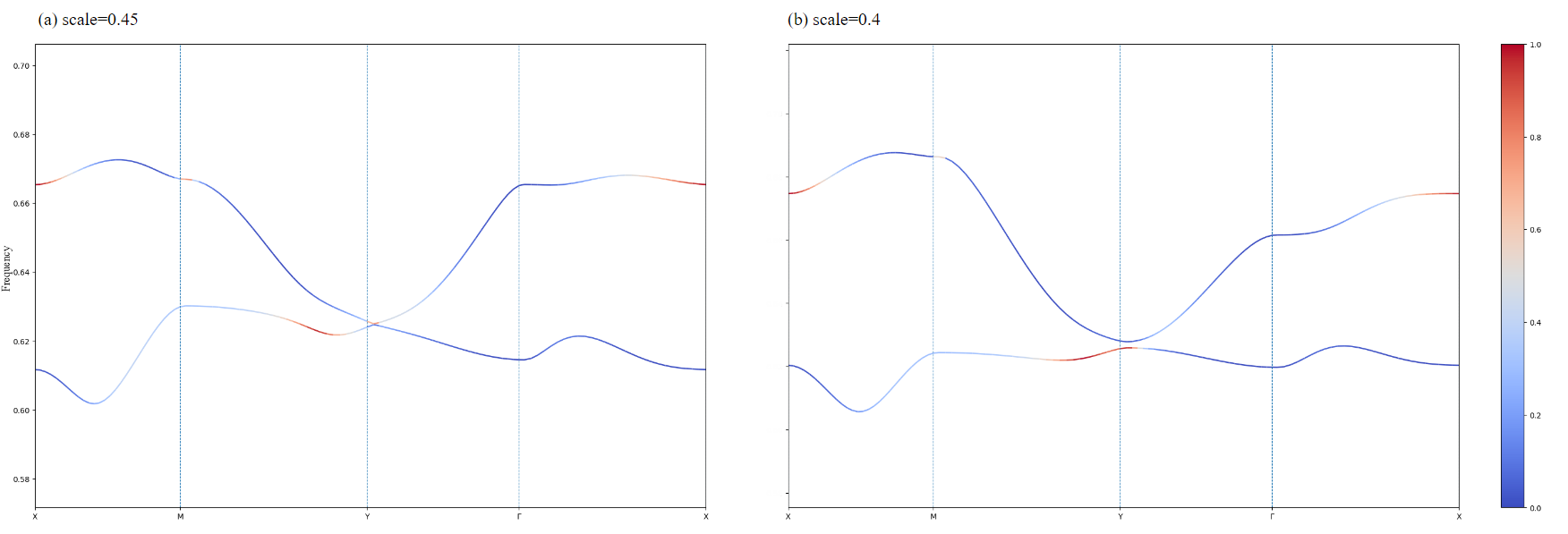}
\caption{C$_2$ square cell (FD–Bloch, TM). Bands along $X\to M\to Y\to\Gamma\to X$ for two dimerization scales: (a) $\text{scale}=0.45$, (b) $\text{scale}=0.40$. Two identical Si pillars sit on the anti-diagonal of the square cell. The parameter \emph{scale} multiplies the pillar coordinates measured from the cell center; smaller values pull the pillars inward and strengthen intra-cell hybridization. Colors indicate the normalized $p$-subspace weight (blue $p$-poor, red $p$-rich). Panel (a) approaches a near-degenerate pair at $Y$ without inversion. Panel (b) shows an exchange of orbital character \emph{at $Y$} and a linear reopening away from $Y$, which diagnoses a $Y$-point inversion.}
\label{fig:C2}
\end{figure*}
We place two identical pillars on the anti-diagonal of a square Bravais cell and tune a dimensionless dimerization parameter \(\mathrm{scale}\). We set \(\mathbf a_1=(a,0)\), \(\mathbf a_2=(0,a)\) and \(\mathbf b_1=(2\pi/a,0)\), \(\mathbf b_2=(0,2\pi/a)\); the high-symmetry points are \(\Gamma=(0,0)\), \(X=(\pi/a,0)\), \(Y=(0,\pi/a)\), \(M=(\pi/a,\pi/a)\). We follow the path \(X\!\to\!M\!\to\!Y\!\to\!\Gamma\!\to\!X\) as in Fig.~\ref{fig:C2}.

We expand about the Y point with local crystal momentum \(\mathbf q=\mathbf k-\mathbf k_Y\). The little group at \(Y\) (isomorphic to \(C_{2v}\)) admits linear inter-orbital coupling, so we write the minimal block as $ h_Y(\mathbf q)=\big(M+\mathbf q^{\mathsf T}B\,\mathbf q\big)\,\tau_z
+ v_x\,q_x\,\tau_x + v_y\,q_y\,\tau_y ,$ where \(q_x\) points along \(Y\!\to\!M\) and \(q_y\) along \(Y\!\to\!\Gamma\); \(B\) is a real symmetric curvature tensor and \(v_{x,y}\) are real velocities capturing the axial anisotropy. We write the band energies in compact form as $\omega_\pm(\mathbf q)=d_0(\mathbf q)\pm\sqrt{(M+\mathbf q^{\mathsf T}B\,\mathbf q)^2+v_x^2 q_x^2+v_y^2 q_y^2}\,.$
where we present the expression inline to highlight the linear gap reopening governed by \(v_{x,y}\).

Fig.~\ref{fig:C2}(a) (\(scale=0.45\)) brings a near-degenerate partner to \(Y\) while keeping the ordering; the colored weights vary smoothly along \(M\!-\!Y\) and \(Y\!-\!\Gamma\). Reducing the scale to \(0.40\) [Fig.~\ref{fig:C2}(b)] drives the mass \(M\) through zero at \(Y\): the two branches exchange orbital character at \(Y\) and the gap reopens linearly for small \(|\mathbf q|\). The unequal slopes along \(Y\!-\!\Gamma\) and \(Y\!-\!M\) confirm \(v_y\neq v_x\) and therefore the anisotropic linear coupling expected for \(C_2\).

\textit{Conclusion.-}We have established a unified symmetry-constrained $k \cdot p$ framework that accurately models TM photonic bands near high-symmetry points through two low-energy subspaces and a partnered doublet, formulated as $h_\eta(\mathbf{k}) = d_0 \tau_0 + d_z \tau_z + \big[ \Phi_\eta \tau_+ + \Phi_\eta^* \tau_- \big]$. This approach successfully captures the distinct behavior across four crystal symmetries: in $C_6$ systems, intra-cell contraction drives the mass term $M$ through zero at $\Gamma$ with linear reopening governed by $m = 1$ selection rules; in $C_4$ crystals, preserved mirror symmetries enforce quadratic $m = 2$ coupling, producing characteristic mirror-aligned anisotropy patterns; for $C_3$ symmetry, the absence of a second partner doublet prevents $\Gamma$-point inversion despite available linear coupling channels; and in $C_2$ systems, anti-diagonal dimerization induces $Y$-point inversion with anisotropic linear reopening. Throughout our investigation, we have demonstrated the capability to directly extract crucial parameters—including the mass term $M$, minimal coupling order $m$, and anisotropy—from projectors, dispersions, and Brillouin zone maps without coefficient fitting. This symmetry-first framework provides a powerful and versatile tool for analyzing, comparing, and designing band inversions and coupling orders across $C_6$, $C_4$, $C_3$, and $C_2$ photonic crystals, offering significant potential for advanced topological photonic engineering.  

\textit{Declaration of competing interest.-} The authors declared that they have no conflicts of interest to this work.
\textit{Acknowledgment.-}This work is supported by the Developing Project of Science and Technology of Jilin Province (20240402042GH). 

\textit{Data availability.-}Data will be made available on request.

\appendix

\bibliography{apssamp}

\end{document}